\documentclass[prd,reprint,nofootinbib,notitlepage,aps,tightenlines,amsmath,amssymb,showpacs,superscriptaddress]{revtex4-1}

\usepackage[hyperindex,breaklinks]{hyperref}
\usepackage{graphicx,epstopdf}

\usepackage{tikz}
\usetikzlibrary{shapes,arrows}

\usepackage{natbib,bm}
\usepackage{slashed}    
\usepackage{hyperref}
\usepackage{breakurl}
\usepackage{multirow}
\usepackage{bbold}
\usepackage{listings}

\def\be{\begin{equation}}
\def\ee{\end{equation}}
\def\bea{\begin{eqnarray}}
\def\eea{\end{eqnarray}}
\def\ge{\mathrel{\raise.3ex\hbox{$>$\kern-.75em\lower1ex\hbox{$\sim$}}}}
\def\la{\mathrel{\raise.3ex\hbox{$<$\kern-.75em\lower1ex\hbox{$\sim$}}}}

\def\simgt{\mathrel{\raise.3ex\hbox{$>$\kern-.75em\lower1ex\hbox{$\sim$}}}}
\def\simlt{\mathrel{\raise.3ex\hbox{$<$\kern-.75em\lower1ex\hbox{$\sim$}}}}

\newcommand{\nc}{\newcommand}

\nc{\gone}{\bar g_{\pi NN}^{(1)}}
\nc{\gzero}{\bar g_{\pi NN}^{(0)}}
\nc{\al}{\alpha}
\nc{\ga}{\gamma}
\nc{\de}{\delta}
\nc{\ep}{\epsilon}
\nc{\ze}{\zeta}
\nc{\et}{\eta}
\nc{\ka}{\kappa}
\nc{\rh}{\rho}
\nc{\si}{\sigma}
\nc{\ta}{\tau}
\nc{\up}{\upsilon}
\nc{\ph}{\phi}
\nc{\ch}{\chi}
\nc{\ps}{\psi}
\nc{\om}{\omega}
\nc{\Ga}{\Gamma}
\nc{\De}{\Delta}
\nc{\La}{\Lambda}
\nc{\Si}{\Sigma}
\nc{\Up}{\Upsilon}
\nc{\Ph}{\Phi}
\nc{\Ps}{\Psi}
\nc{\Om}{\Omega}
\nc{\ptl}{\partial}
\nc{\del}{\nabla}
\nc{\ov}{\overline}
\nc{\newcaption}[1]{\centerline{\parbox{15cm}{\caption{#1}}}}
\nc{\us}{U(1)$_S$}

\mathchardef\mhyphen="2D

\def\beq{\begin{equation}}
\def\eeq{\end{equation}}
\def\bmat{\begin{displaymath}}
\def\emat{\end{displaymath}}
\def\bear{\begin{eqnarray}}
\def\eear{\end{eqnarray}}
\def\ba{\begin{eqnarray}}
\def\ea{\end{eqnarray}}
\def\bery{\begin{array}}
\def\ery{\end{array}}
\def\bit{\begin{itemize}}
\def\eit{\end{itemize}}
\def\ben{\begin{enumerate}}
\def\een{\end{enumerate}}
\def\btab{\begin{tabular}}
\def\etab{\end{tabular}}
\def\btbl{\begin{table}}
\def\etbl{\end{table}}
\def\bfig{\begin{figure}[htb]}
\def\efig{\end{figure}}
\def\bpic{\begin{picture}}
\def\epic{\end{picture}}

\def\ga{\mathrel{\raise.3ex\hbox{$>$\kern-.75em\lower1ex\hbox{$\sim$}}}}
\def\la{\mathrel{\raise.3ex\hbox{$<$\kern-.75em\lower1ex\hbox{$\sim$}}}}
\def\gappeq{\mathrel{\rlap {\raise.5ex\hbox{$>$}}
{\lower.5ex\hbox{$\sim$}}}}
\def\lappeq{\mathrel{\rlap{\raise.5ex\hbox{$<$}}
{\lower.5ex\hbox{$\sim$}}}}

\def\gyr{{\rm \, G\kern-0.125em yr}}
\def\mev{{\rm \, Me\kern-0.125em V}}
\def\gev{{\rm \, Ge\kern-0.125em V}}
\def\tev{{\rm \, Te\kern-0.125em V}}

\newcommand{\keV}{\ensuremath{\mathrm{keV}}}
\newcommand{\MeV}{\ensuremath{\mathrm{MeV}}}
\newcommand{\GeV}{\ensuremath{\mathrm{GeV}}}

\renewcommand{\sec}{\ensuremath{\mathrm{s}}}

\newcommand{\cm}{\ensuremath{\mathrm{cm}}}
\newcommand{\Mpc}{\ensuremath{\mathrm{Mpc}}}

\newcommand{\Gyr}{\ensuremath{\mathrm{Gyr}}}

\begin{document}

\title{Signatures of Dark Radiation in Neutrino and Dark Matter Detectors}

\author{Yanou Cui}
\affiliation{Department of Physics and Astronomy, University of California, Riverside, CA 92521, USA}

\author{Maxim Pospelov}
\affiliation{Perimeter Institute for Theoretical Physics, Waterloo, ON N2J 2W9, 
Canada}
\affiliation{Department of Physics and Astronomy, University of Victoria, 
  Victoria, BC V8P 5C2, Canada}
\affiliation{CERN, Theoretical Physics Department, Geneva, Switzerland}

\author{Josef Pradler}
\affiliation{Institute of High Energy Physics, Austrian
     Academy of Sciences, Nikolsdorfergasse 18, 1050 Vienna,
     Austria}

\begin{abstract}
\noindent 
We consider the generic possibility that the Universe's energy budget
includes some form of relativistic or semi-relativistic dark radiation
(DR) with non-gravitational interactions with Standard Model (SM)
particles.  Such dark radiation may consist of SM singlets or a
non-thermal, energetic component of neutrinos. If such DR is created
at a relatively recent epoch, it can carry sufficient energy to leave
a detectable imprint in experiments designed to search for very weakly
interacting particles: dark matter and underground neutrino
experiments. We analyze this possibility in some generality, assuming
that the interactive dark radiation is sourced by late decays of an
unstable particle, potentially a component of dark matter, and
considering a variety of possible interactions between the dark
radiation and SM particles. Concentrating on the sub-GeV energy
region, we derive constraints on different forms of DR using the
results of the most sensitive neutrino and dark matter direct
detection experiments.  In particular, for interacting dark radiation
carrying a typical momentum of $\sim30$~MeV$/c$, both types of
experiments provide competitive constraints.  This study also
demonstrates that non-standard sources of neutrino emission ({\em
  e.g.} via dark matter decay) are capable of creating a ``neutrino
floor" for dark matter direct detection that is 
closer to current bounds than is expected from standard neutrino
sources.

\end{abstract}
\maketitle

\section{Introduction}

The dominance of dark matter (DM) and dark energy (DE) in the total
energy balance of the Universe is a widely acknowledged 
nonetheless astounding fact. Precision studies of the cosmic
microwave background (CMB) \cite{Ade:2015xua} provide evidence that DM
was ``in place" before hydrogen recombination, while the effects of DE
manifest themselves much later, starting from redshift $z\sim O(1)$
\cite{Weinberg:2008zzc}. Given a rather elaborate structure of the
visible sector, the Standard Model of particles and fields (SM), it
would be a logical imperative to consider somewhat more complicated
models of a dark sector, beyond a single particle contributing to DM,
and a cosmological constant sourcing  DE. Indeed, in recent years
there has been some significant progress in studying models of dark
sectors \cite{Essig:2013lka}, which  include the possibility of new
``dark forces", new massless states (also known as ``dark radiation",
or DR), and/or relativistic massive dark states that can be produced
through late time processes (also known as ``boosted dark matter'')
\cite{Huang:2013xfa,Agashe:2014yua, Berger:2014sqa, Kachulis:2017nci}, co-existing with massive DM particles. Both new
massless states and massive boosted states in the dark sector can be
classified as ``dark radiation'' in general terms.  The presence of
such sectors significantly broadens the phenomenology of DM
\cite{Boehm:2003hm, Pospelov:2007mp, ArkaniHamed:2008qn,
  Jaeckel:2010ni, Agashe:2014yua, Berger:2014sqa, Kong:2014mia, Bhattacharya:2014yha, Agashe:2015xkj,
  Alhazmi:2016qcs, Necib:2016aez, Cherry:2015oca, Kopp:2015bfa, Hu:2016xas, Kim:2016zjx, Bhattacharya:2016tma}
motivating, in turn, a wider scope for the experimental efforts
dedicated to the searches of DM.

An enormous progress in observational cosmology has resulted in a very
sensitive constraint on the number of extra degrees of freedoms that
remained radiation-like during the CMB epoch. The Planck collaboration
has reported a stringent constraint, that phrased in terms of neutrino-like
radiation species reads as \cite{Ade:2015xua}: \be N_{\rm eff}
=3.04\pm 0.33 ~~\Longrightarrow ~~\rho_{\rm DR}/\rho_\gamma < 0.15 ,
\label{Neff}
\ee where $\rho_{\rm DR}$ is the energy density in additional dark
radiation. With further refinements \cite{Baumann:2015rya, Brust:2017nmv}, one can show
that the actual $2\sigma$ limit on the deviation from the SM
prediction for $N_{\rm eff}$ is $\Delta N_{\rm eff}\leq 0.39$.  This
bound has a wide degree of applicability, and is most effectively used
to constrain models with ``early" DR, or models with extra light
degrees of freedom that were in thermal contact with the SM, but
decoupled at some point in the history of the early Universe. In
that case there are also extra bounds provided by big bang
nucleosynthesis (BBN), that can be often cast in terms of the same
parameter $N_{\rm eff}$~\cite{Nollett:2013pwa,Nollett:2014lwa,Boehm:2013jpa}.  However, constraint (\ref{Neff}) would not
be applicable to models where DR is created much later than the CMB
epoch.  For example, recent decays of a sizable fraction of DM into
dark radiation are allowed, and, moreover, $\rho_{\rm DR}$ can be much
larger than $\rho_\gamma$ today.

In this paper, we are interested in the late generation of
$\rho_{\rm DR}$ with the following properties: the number density of
DR particles is smaller than that of CMB photons, while the kinetic
energy on average is much larger than the energy of individual CMB
quanta, \be n_{\rm DR} \ll n_\gamma,~~ E_{\rm DR} \gg
E_{\gamma},~~\rho_{\rm DR} (\sim E_{\rm DR}n_{\rm DR})\leq 0.1
\rho_{\rm DM}
\label{ineq}
\ee In the last relation, we require that the amount of dark radiation
does not exceed 10\% of the dark matter energy density, in accordance
with recently updated constraints \cite{Poulin:2016nat}. Such set of
inequalities leaves, of course, a lot of freedom for what DR can be,
but restricts a number of possibilities for how the non-thermal DR can
be created.

In this paper, we will consider a scenario where the dark sector
mediates some DR-SM interaction to be specified below.  The new
interactions allow to probe DR directly via its interaction with
nuclei and electrons rather than via its contribution to the
Universe's energy balance (through $\rho_{\rm DR}$.)  Cosmic SM
neutrinos are, in some sense, the example of interacting dark
radiation. Remnants from the Big Bang, they have a very small energy
today $E_\nu \sim m_\nu$, and their direct detection via weak
interactions represents a huge experimental
challenge~\cite{Betts:2013uya}. (There is, of course, plenty of
evidence for cosmic neutrino weak interactions in the outcome of BBN.)
It is also known that neutrinos have other cosmic components,
including the one generated by global activity of supernova (SN)
remnants~\cite{Beacom:2010kk}. The search for the diffuse SN neutrino
flux is a challenging endeavor, driving some developments in solar
neutrino detectors~\cite{Beacom:2003nk}.

Neutrinos provide a small---but in future important---background for
the searches of weakly interacting massive particles (WIMPs) in direct
detection experiments
\cite{Monroe:2007xp,Strigari:2009bq,Drukier:1983gj}.%
\footnote{The idea to study the coherent scattering of neutrinos on
  nuclei~\cite{PhysRevD.9.1389} pre-dates the DM direct detection idea~\cite{Goodman:1984dc},
  and only recently has it been observed with neutrinos sourced by
  meson decays~\cite{Akimov:2017ade}.}  The solar neutrino flux is the
largest component for such a background, but given a relatively low
energy cutoff to its spectrum, the generated nuclear recoil is rather
soft. Above a neutrino energy of 18~MeV only diffuse supernova and
atmospheric neutrinos constitute the primary components of the
neutrino spectrum. It has to be emphasized that there is no direct
observation of the diffuse supernova flux yet, and only an upper limit
exists on the flux of electron antineutrinos in the 15-30~MeV window
\cite{Bays:2011si}, $\Phi_{\bar\nu_e} < 3/\cm^2/\sec$. Above
  30~MeV, the atmospheric neutrinos start to be dominant, with
  measurements above 150~MeV energy~\cite{Richard:2015aua}, and a
  total flux in the ballpark of a ${\rm few}/\cm^2/\sec $.  The above
window is quite important because it corresponds to a momentum
transfer scale that is associated with the optimum sensitivity region
of dark matter detectors, such as large xenon-based detectors
\cite{Akerib:2015rjg,Aprile:2012nq,Tan:2016diz}. Therefore, \textit{if} 
 there is an additional neutrino or neutrino-like component of dark
radiation, the bottom of the direct detection ``neutrino floor" can be
closer than expected.%
\footnote{An upper limit on a cosmological flux of SM neutrinos from
  DM decay in the $\sim$15-100~MeV mass window using Super-Kamiokande
  data has previously been established in~\cite{PalomaresRuiz:2007ry};
  see also recent Ref.~\cite{Garcia-Cely:2017oco}. The ensuing lower
  limit on the DM lifetime is driven by the $\bar \nu_e$
  flux-component; below we will revisit this limit, by assuming the
  absence of anti-neutrinos in the decay of DM. For limits on the
  $\nu +\bar \nu$ flux from MeV-mass DM annihilation
  see~\cite{PalomaresRuiz:2007eu}.}
Given the huge amount of efforts devoted to the
scaling-up of the WIMP direct detection experiments, one should
investigate possible signals from additional hypothetical components
of the neutrino flux, and from DR in general. Apart from creating a
new signal competing with WIMPs, the search for DR is interesting in
its own right, and can be done in parallel to WIMP searches. It is
well known that direct detection experiments often provide the most
sensitive tool for broad classes of physics with 1~eV to a few 100~keV
energy release (see, {\em e.g.},
Refs.~\cite{Pospelov:2008jk,Graham:2012su,Essig:2011nj,An:2013yua,An:2014twa,Hochberg:2015pha,Essig:2015cda}
for a sample of representative ideas), and therefore it is easy to
anticipate that they can be used as tools for exploring DR.

In this paper we address several questions related to DR, WIMP direct
detection experiments, and neutrino physics. Our primary goal is to
assess---in broad classes of models---the sensitivity reach of direct
detection experiments to DR. We shall concentrate on the eV-GeV
energy range, and will mostly assume that a certain fraction of DM is
prone to decay to DR. In the next section, we will outline several
classes of possible interactions. In Sec.~\ref{minF} we provide the
computation of DR fluxes resulting from decaying DM, both in our halo
and globally in the Universe.  Section \ref{DRsignal} considers the main
phenomenological features of DR in several classes of generic models:
DR in the form of the new population of neutrinos or neutrino-like light fermions, axion-like
particles and light vectors (``dark photons"). We provide further
discussion and conclusions in Section \ref{DC}.

\section{DR-SM interactions}

Among the possible sources of DR there can be processes involving the
collision and decay of SM particles, and, under certain conditions,
collisions, annihilation, and/or decay of massive DM particles. It is
apparent that SM processes cannot create large amounts of DR, and in
particular cannot saturate the last inequality in
Eq.~(\ref{ineq}). The reason is that the only steady source of DR
emitted globally in the Universe can come either from cosmic rays
collisions with interstellar medium, or from the production of DR
inside stars, including SN explosions. The energetics of these
processes is subdominant to the energy locked inside $\rho_{\rm DM}$ by many
orders of magnitude.

On the contrary, very limited amount of information about the
properties of DM provides some grounds for speculation that DM can
indeed be a powerful source of DR. In the rest of this paper, we will
concentrate on the decays of DM progenitor particles, giving rise to
DR.  We will call such a DM progenitor particle as $X$, and let
$\kappa$ be an (energy) fraction of DM that is allowed to decay.
There are two generic types of DR we may consider. First, the DR can
be the SM neutrinos: DM decay provides a late time source of energetic
$\nu$ injection, and the SM-DR interaction in this case are the
familiar neutrino weak interactions. Second, the quantum of DR could
be a SM singlet, which we will label as $\chi$; the pattern of the SM-$\chi$
interaction is more uncertain and diverse in this case. In the
following we will discuss the general logical possibilities for these
two cases.

\subsection{SM neutrinos as dark radiation}
In this scenario, we consider the possibility that DM gives rise to
DR in the form of SM neutrinos. This can occur directly by
$X\to \nu(\bar \nu)+Y$ decays (where $Y$ stands for the rest of the
decay products).  This can also occur in two steps: first through the
decay of $X$ to a nearly massless fermion $\chi$ ({\em e.g.}, a sterile
neutrino), that then oscillates into the SM neutrinos under certain
conditions.  Models with direct decay to neutrinos are free from
potential constraints from $N_{\rm eff}$ measurements,  the $\nu$-SM
interactions are known, and the model is more minimal/simple. Then, on
top of the constraints from dark matter experiments that limit the
neutral current interactions of the DR neutrinos with nuclei, there
will be additional constraints provided through weak
interactions at neutrino detectors such as Super-Kamiokande (SK).  The scenario with
decays to neutrinos is also interesting in that it can be motivated by
certain neutrino mass generation mechanisms, and connects to other
aspects of neutrino physics and observables.  Possibilities include: a
vector or scalar boson DM particle $X$ decaying to a pair of $\nu$
($\bar{\nu}$), or fermionic DM $X$ decaying to $\nu+Y$ where $Y$ is a
vector/scalar boson. Here we focus on the representative case where 
scalar DM $X=\phi$ decays to $\nu\nu$ and/or $\bar{\nu}\bar{\nu}$.

\subsubsection{Goldstone boson Majoron DM decaying to neutrinos}
It has been considered that neutrino masses may arise from a global
symmetry breaking (see e.g. \cite{Chikashige:1980ui,Schechter:1981cv}
as well as \cite{Chacko:2003dt} and references therein). The effective
coupling of the Goldstone boson (Majoron) $\phi$ to neutrinos, resulting from
the breaking of this global symmetry, is given by
 \be
\mathcal{L}\supset \frac{\phi H^2L_iL_j}{\Lambda_{ij}^2}\rightarrow
i\frac{1}{2}g_{ij}\phi\nu_i\nu_j + h.c. \label{eq:majoron_nu}, \ee 
where $L_i$ stands for the SM lepton doublets, $H$ is the SM Higgs boson, and $\Lambda_{ij}$ roughly corresponds to
the global symmetry breaking scale. We have
inserted the Higgs vacuum expectation value (vev) at the second
step in the above equation. Note that here we retained the most generic flavor pattern in
the coupling $g_{ij}$ which can account for neutrino mass and
mixing. For simplification, we suppress the flavor indices of neutrinos and assume flavor-diagonal couplings in later discussion. 
In this case $\phi$ is real, and it decays to $\nu,\bar{\nu}$
symmetrically.

The UV completion of the effective interaction in equation
(\ref{eq:majoron_nu}) can be the original Majoron model
\cite{Chikashige:1980ui} where the authors assume a singlet Higgs
field $\Phi$ responsible for the lepton number ($L$) symmetry
breaking, and a Dirac pair of singlet fermions $S_L, S_R$ which give
rise to the familiar see-saw right-handed neutrino $N$ once
$L$-symmetry is broken. The Lagrangian relevant to $m_\nu$-generation
is (we drop the lepton flavor indices)
\be
\mathcal{L}=y_1\bar{L}HS_R+y_2\Phi\bar{S}_LS_R+h.c.
\ee
Upon symmetry breakings, this leads to two non-vanishing elements in the mass matrix of the system: $m=y_1\langle H\rangle$, $M=y_2\langle \Phi\rangle$.
Integrating out the heavy singlet fermions after symmetry breaking, one finds the prediction for neutrino masses and the Majoron-$\nu$ couplings as (assuming $m\ll M$):
\bea
m_\nu&=& m^2/M=\frac{y_1^2\langle H\rangle^2}{y_2\langle \Phi\rangle}\\
\mathcal{L}_{\phi\nu\nu}&=&i\frac{m_\nu^2}{\langle H\rangle^2}\frac{y_2}{y_1^2}(\nu\nu-\bar{\nu}\bar{\nu})\phi.
\eea

We can readily see how the parameters in the UV complete theory map to the effective coupling $g$ in Eq. (\ref{eq:majoron_nu}): $g\simeq\frac{y_2m_\nu^2}{y_1^2\langle H\rangle^2}$.
The mass of a Majoron as a pseudo-Goldstone has large uncertainty,
generated by additional amounts of explicit global symmetry breaking
({\em e.g.} via a possible breaking of global symmetry by quantum
gravity). There have been studies on the possibility of Majoron DM, and the focus has
been on thermally produced Majoron, that typically needs to have mass O(keV) or
less in order not to over-close the
Universe~\cite{Berezinsky:1993fm, Lattanzi:2007ux, Esteves:1276456, Gu:2010ys, Lattanzi:2014mia, Garcia-Cely:2017oco}. A Majoron with mass O(10)~MeV or
higher (which is our interest) can be relieved from this cosmological
bound if it is produced non-thermally, through late decay of a massive
particle (\textit{e.g.} modulus) or late during inflation/reheating.
One can check that realistic neutrino masses and a Majoron with
lifetime around the age  of the Universe can be simultaneously accommodated in
this model, with perturbative couplings and Majoron masses in the
range of our interest. In particular, the decay rate of $\phi$ in terms of effective parameter $g$ or UV parameters can be written as follows:
\bea
&&\Gamma_\phi\simeq t_0^{-1}\left( \frac{m_\phi}{20~\rm MeV} \right)\left( \frac{g}{6\times 10^{-20}} \right)^2\\
&\simeq&t_0^{-1}\left( \frac{m_\phi}{20\rm~ MeV} \right)\left( \frac{\sum_i m_{\nu_i}^2}{(0.1~\rm eV)^2} \right)\left( \frac{\langle\Phi\rangle}{3\times 10^9~\rm GeV} \right)^{-2}
\nonumber
\eea
where $t_0=4.3\cdot10^{17}\,\rm s$, the age of the universe, has been factored out for 
convenience. 

A dedicated study of such a decaying Majoron DM
and its relation to neutrino physics is an interesting topic on its
own, but falls outside the scope of this
paper. 

\subsubsection{Complex Scalar boson DM decaying to neutrinos}
\label{sec:complex-scalar-boson}

This scenario is inspired by the above Majoron model, yet instead we
consider a complex scalar that carries lepton number but does not
condense. Now the relevant interaction is: 
\be \mathcal{L}'\supset
\frac{1}{2}g\phi'\bar{\nu}\bar{\nu} + \frac{1}{2}g{\phi'}^{*}\nu\nu . 
 \ee 
 The UV completion of
this model is similar to the Majoron model. But instead of $\Phi$ in the earlier model, we introduce a complex inert singlet, $\phi'$, which carries
$L$-number but does not condense like $\Phi$. Notice that the light
scalar $\phi'$ is not a Goldstone boson, so that its mass is not protected against radiative corrections. 
Thus, a sub-electroweak scale mass of $\phi'$ may be
associated with its own naturalness problem, which we do not address
here. The relevant Lagrangian in this case has a
similar form to that of the earlier model involving $\Phi$: \be
\mathcal{L}'=y'_1\bar{L} HS_R+y'_2\phi'\bar{S}_LS_R+h.c.  \ee After the $L$ -symmetry
breaking, {\em e.g.}~by $\Phi$ condensation, and after integrating out heavy states, we
find: \be \mathcal{L}_{\phi'\nu\nu}=\frac{m_\nu^2}{\langle
  H\rangle^2}\frac{y'_2}{{y'_1}^2}({\phi'}^{*}\nu\nu+\phi'\bar{\nu}\bar{\nu}).  \ee

This model allows the possibility that there is an asymmetric
abundance of $\phi'$ vs. ${\phi'}^{*}$, so that neutrinos instead of
anti-neutrinos are produced as DR. Consequently, DR composed of $\nu$ rather than $\bar\nu$ 
may not be subject to strong constraints imposed by SK on the $\bar{\nu}_e$ flux just above the endpoint of 
the solar neutrino spectrum. The asymmetric DM type of $\phi'$ requires an
initial asymmetry generation (\textit{e.g.}~through the decay of a
heavy Majorona RH neutrino) and efficient symmetric annihilation that
depletes ${\phi'}^{*}$. It is possible to realize such a model by extending the  ingredients to include
additional light (or massless) states enabling $\phi'{\phi'}^{*}$ annihilation. 

\subsubsection{DM decay giving rise to SM neutrinos through mass mixing}
Another possibility of DM $X$ decaying to SM neutrinos is through the
mixing between SM neutrinos and light singlet sterile neutrinos that
directly couple to DM. This falls into a similar category, as we will
discuss in Sec.~\ref{sec: singlet_DR}: $X \to \chi +\chi$ combined
with the linear operator of mass mixing
$m_{\chi\nu}\chi \nu_{\rm SM}$. Depending on the $\chi-\nu$
oscillation parameters, $\chi$ may or may not contribute to
$N_{\rm eff}$. For example if the effective oscillation length between
$\chi$ and $\nu$ is astronomically large, which can be achieved for
nearly mass degenerate states $\chi$ and $\nu$ (and therefore very
light $\chi$), $N_{\rm eff}$ does not provide an immediate constraint,
as the $\nu \to \chi$ oscillation rate in the early Universe can be
arbitrarily small. If in such a model the decays produce $\chi$'s of
certain helicity, then the oscillations may result in the predominance
of $\nu$ over~$\bar\nu$.

\subsection{SM singlets as dark radiation}\label{sec: singlet_DR}

One can have several logical possibilities in this case: \be X \to
\chi +\chi,~~{\rm or} ~~ X \to Y+ \chi,~~{\rm or} ~~ X \to {\rm SM}
+\chi~~{\rm etc.}  \ee In general, such processes may occur with or
without SM particles, or other members of dark sector particles $Y$ in
the final state, and the multiplicity of $\chi$ may vary. Also, the DR
particles can be fermions or bosons. In other words, at this point
there appears to be a great freedom in choosing models for $X$ and
$\chi$, apart from $\gamma$- and cosmic-ray constraints if the decay
is accompanied by energetic SM particles.

\subsubsection{Singlet $\chi$ as a fermion: scattering signal}
Let us assume, for a moment, that DR $\chi$ is a fermion. Then,
quite generically, the most important interactions of $\chi$ with the
SM occur either at the linear or bi-linear order in $\chi$, \be {\cal
  L}_{\rm SM- \chi} = (\chi \times O^{\rm SM}_f + h.c.) + (\bar\chi
\Gamma \chi) \times O^{\rm SM}_b, \ee where $O^{\rm SM}_{f(b)} $ are
generic fermionic or bosonic composite operators built from the SM
fields; $(\bar{\chi} \Gamma \chi) $ stands for a generic bilinear operator
in $\chi$, and may contain a variety of currents, such as
$\bar \chi \gamma_\mu\chi$, $\bar\chi \chi$ etc.  The overall
transformation properties of $O^{\rm SM}_{f(b)} $ must be chosen such
as to make these terms in the Lagrangian Lorentz invariant. To narrow
down these many possibilities, we will consider \be O^{\rm SM}_f =
m_{\chi\nu} \nu_{\rm SM},
\label{osc}
\ee where $\nu_{\rm SM}$ is some linear combination of  SM
neutrinos, and $m_{\chi\nu}$ is the effective mass parameter that will
mix SM neutrinos and $\chi$ fermions. This mass parameter can be
thought of as the low-energy limit of an operator respecting all
symmetries of the SM. This way, $\chi$ will ``participate" in the
neutrino mass and mixing matrices.

Among the bilinear interactions we will consider a subset of the most
relevant ones for direct detection phenomenology, \be (\bar \chi
\Gamma \chi ) \times O^{\rm SM}_b = (\bar \chi \gamma_\mu \chi )
\times ( G_V J_{EM}^\mu + G_B J_B^\mu ),
\label{GVGB}
\ee with two currents $J_{EM}$ and $J_B$ representing the SM
electromagnetic current (defined without a  factor of $e$), and the
baryonic current of quarks.  At the level of electrons, protons and
neutrons, these currents can be approximated as \be J_{EM}^\nu = -\bar
e \gamma^\mu e + \bar p \gamma^\mu p; ~~ J_{B}^\nu = \bar n \gamma^\mu
n + \bar p \gamma^\mu p, \ee where corrections related to the
anomalous magnetic moments have been neglected. There are well-known
UV completions of these operators where the exchange is mediated
either by the dark photon (a vector particle $V_\mu$ with kinetic
mixing parameter $\epsilon$ to the photon field and 
$Q_\chi g' V_\mu (\bar \chi \gamma_\mu \chi )$ coupling), and/or by the
baryonic vector (a new vector particle $V^{B}_\mu$ coupled to a baryon
current via $g'V^{B}_\nu J_{B}^\nu$).  In that case the effective
couplings $G_V$ and $G_B$ can be expressed in terms of the more
fundamental parameters as 
\be G_V = \frac{ g'\epsilon e Q_{\chi}}{m_V^2};~~G_B
= \frac{ g'^2Q_{\chi}}{m_V^2},
\label{GVGB}
\ee
where $Q_\chi$ is the charge of $\chi$ under the dark $U(1)'$.
While the dark photon models are fully UV
consistent without further modifications, the baryonic model needs to
be augmented at the weak scale to cancel the gauge anomaly in that
sector. In the last decade, the phenomenological aspects of these
models have been considered in some detail
\cite{Pospelov:2007mp,Batell:2009di,Pospelov:2011ha,Izaguirre:2014bca,Pospelov:2013rha,e137,Batell:2014yra,Tulin2014}. Notice
that recent works \cite{Dror:2017ehi,Dror:2017nsg} have significantly
advanced constraints on gauged baryon models, when $V^{B}_\mu$ is
assumed to be light, and the gauge anomalies are canceled by a new set
of fermions above the weak scale. Nevertheless such constraints do not directly apply, as 
(\ref{GVGB}) includes an extra parameter, $Q'_\chi$, that can be large 
without violating perturbative unitarity (that requires $Q_\chi g'$ to be less than $4 \pi$). In this paper, however, we will stay
on phenomenological grounds and treat $G_B$ as a free parameter. 
Also note that if mass of the mediator is zero in the $G_V$ coupling, 
$\chi$ will appear as fractionally charged fermion with effective EM charge of 
$\epsilon Q_\chi g'$. Milli-charged particles is an interesting and viable case of DR, as we will
 briefly discuss later.
 
While the choices given in (\ref{osc}) and (\ref{GVGB}) are far from being
exhaustive, they are better motivated than many other ad-hoc models.
Moreover, they are sufficient to capture the main phenomenological
possibilities: $\nu-\chi$ oscillation transitions and scattering of DR
states on electrons and nuclei.

\subsubsection{Singlet $\chi$ as a boson: absorption signal}
If DR is represented by light bosonic particles,
the possible choices of interactions are again plentiful. Natural
possibilities of light bosons include dark photons and axion-like
particles,  both having masses protected by symmetries. These
possibilities are also well-explored in the context of new directions
for DM searches (see {\em e.g.}  \cite{Battaglieri:2017aum}).  We
first consider the decay of $X$ to two dark photons, $X\to VV$. This
is a very economical model, where the interaction of dark photons with
the SM is governed by the mixing parameter $\epsilon$ and the mass of
the dark photon $m_V$ \cite{Holdom:1985ag}. In vacuum, the dark photon
couples to the electromagnetic current via
$V_\mu \times ( e\epsilon J_{EM}^\nu)$.  In medium, there is a
well-known suppression of this interaction in the regime where $m_V$ is
much smaller than the plasma frequency
\cite{Redondo:2008aa,An:2013yfc,Graham:2014sha}.  The other
possibility of bosonic $\chi$ is an axion-like particle (ALP), $a$. The axion-like $a$ can be produced through a scalar $X$ DM
decay $X\to aa$, then $a$ may interact with SM states via
$aF_{\mu\nu}\tilde{F}_{\mu\nu}$ (with the SM photon) or
$a\bar{e}\gamma_5e$ (with the electron). The main difference from
fermionic DR is that bosons can be completely absorbed and converted
to energy carried by the SM. The absorption of dark photons or
ALP-type DM has been considered before in
\cite{Pospelov:2008jk,An:2014twa, Bloch:2016sjj}, and many features of
these analyses can be generalized to ALP/dark photon DR detection.

\subsection{Astrophysical and cosmological constraints on DR}
In the cases where the DR considered here has a small or vanishing
mass, there are many constraints coming from astrophysical and
cosmological observations that one would need to take into account. Apart from the already
mentioned CMB and BBN constraints, mainly in form of $N_{\rm eff}$, there are strong bounds
imposed by stellar energy loss~\cite{Raffelt:1999tx}, which is particularly constraining for the case of $X\rightarrow aa$, $X\rightarrow VV$.
In the case where the DR actually consists of massive (above O(MeV)) yet energetic particles such as boosted DM, BBN and CMB bounds do not
automatically apply, although there can be model-dependent constraints,
depending on how such massive DR interacts with the SM.
 
In several classes of models, DR that can induce observable effects
for the direct detection and neutrino experiments without violating
any of these indirect bounds. Here we list some of the models that
would clearly pass the indirect constraints:
 \begin{enumerate}
 \item Direct decay to neutrinos, $X\to \nu\nu$ for example, does not
   create any additional photons. Electron-positron pairs can be
   produced via $X\to \nu\nu e^+e^-$ but the rate will be suppressed
   by the weak interaction scale. Specifically, one may estimate that
   ${\rm Br}_{X\to \nu\nu e^+e^-} \leq 10^{-3} G_F^2 m_X^4$, where the
   first factor reflects the additional phase space suppression, and
   the maximum available energy scale ($m_X$) is used to render the
   weak branching fraction dimensionless. Even for $m_{X}=1~\GeV$, this is a
   $\sim 10^{-13}$ level of suppression, which will not lead to
   additional constraints on the model from the global production of
   charged lepton cosmic rays.
 
 \item $X$ decay to $\chi\chi$ with bilinear interactions with the SM,
   Eq.~(\ref{GVGB}). The same logic applies, and the probability of
   production of extra SM particles in a single decay process is
   suppressed by the phase space and in addition by $G_{V(B)}^2m_X^4$.
   (We will consider $G_{V(B)}$ in the ballpark of $G_F$).  Here we
   assume, in addition, that the scale of the mediator mass is larger
   than the mass of the decaying particle, $m_V > m_X$, as otherwise a
   much more enhanced on-shell production of mediators may take
   place. With this extra condition, the indirect astrophysical
   constraints can also be completely avoided for sub-GeV~$X$.
 
   These models are, however, constrained by $N_{\rm eff}$. 
   The weak-strength ($G_V\sim G_F$) dark photon mediated interactions are
   capable to keep $\chi$ in a close thermal contact with electrons
   down to $O(\rm MeV)$ temperatures, creating a large positive
   contribution to $N_{\rm eff}$ from the thermal bath of
   $\chi$. Therefore, these models are disfavored unless
   $G_V \ll G_F$, and then consequences of DR on direct detection are
   small. Models with $\chi$ coupled to the baryonic current fare
   better, as hadron number densities diminish rather quickly after
   the QCD phases transition. In this case, the meson-$\chi$
   interaction could decouple relatively early, and $G_B$ on the order
   of $G_F$ is not manifestly excluded.  For that reason we will
   concentrate on models with $G_B$ interaction, noting that the
   thermalization of new degrees of freedom via the baryon current
   interaction is a problem that deserves special separate study. 
   
   In some cases the indirect bounds from $N_{\rm eff}$ can be circumvented at the expense of additional 
   complication of models. For example, if the mass of $\chi$ is in the few MeV range, and, in addition, there is  
   a light mediator particle in the same mass range, $V$, the self-annihilation $\chi\bar\chi \to VV$ with subsequent 
   decay of $V$'s to the SM states may ensure that by the time of SM neutrino decoupling 
   at MeV temperatures, all $\chi$ and $V$ particles from the dark sector have annihilated and decayed, while 
   $X$ DM survives. 
 
 \item $X$ decay to $\chi\chi$, and the transfer of $\chi$ to
   neutrinos via $\chi\to\nu$ oscillations (a possibility that may be called ``neutrino oscillation portal")
   can be made completely safe
   from cosmological and astrophysical bounds as $m_{\chi\nu}$ can be
   almost arbitrarily small. The smallness of this parameter does not
   imply a smallness of the $\chi\nu$ mixing angle with the lowest
   neutrino mass eigenstate in the direct neutrino mass hierarchy
   picture, as there is no lower bound for the lightest
   $m_\nu$. 
   
   \item  Finally, decays of $X$ to nearly massless ALPs $a$, dark photons $V$, 
   and light millicharged particles $\chi$ are not expected to be accompanied by a significant 
   production of the SM states because of the extreme smallness of couplings between SM states 
   and these types of dark sector particles.
 
    \end{enumerate}

\section{Maximum fluxes of DR from decaying DM} 
\label{minF}

Any rate of detection of the relativistic background species $\chi$ (in this section $\chi$ stands for any DR particle, neutrino or singlet $\chi$ as
 discussed earlier) is
controlled by the energy differential particle flux arriving at earth,
$d\phi/dE_{\chi}$. We consider two principal components of this flux,
originating from DM decay within the galaxy (``gal'') or from
extragalactic distances (``eg'').%
\footnote{We neglect small contributions from the local group and
  other  bound structures at cosmological distances.}

We will assume that only a fraction $\kappa$ of the DM density decays (to be consistent with cosmological constraints given in \cite{Poulin:2016nat}),
$\kappa \equiv \left. {\rho_X(t) } /(\Omega_{\rm dm}
    \rho_c)\right|_{t\ll \tau_X}$, where $\tau_X = \Gamma_X^{-1}$ is
the lifetime of the progenitor~$X$. 
Here, $\Omega_{\rm dm} = 0.1198 h^{-2}$ is the CMB-inferred DM density
parameter, and $\rho_c = 3 M_P^2 H_0^2 $ is the critical density
today; $H_0 = 100 h \, km/\sec/\Mpc$ with
$h = 0.6727$~\cite{Ade:2015xua}.%
\footnote{A low-$z$ determination of the Hubble constant gives
  $h=0.73$~\cite{Riess:2016jrr}. It has been suggested that decaying
  DM may alleviate the 2-3$\sigma$ tension between the two inferred values of
  $H_0$~\cite{Berezhiani:2015yta, Chudaykin:2016yfk}. Here we will not
  enter this discussion; recent comments on the viability of this
  possibility are found in~\cite{Poulin:2016nat}.} Clearly, the fluxes
attain their maximum when $\tau_X $ is comparable to the age of the
Universe, $t_0= 13.7\,\Gyr$.

The decay of $X$ injects particles $\chi$ with a spectrum,
\begin{align}
  \frac{dN}{dE_{\chi}} =   \frac{N_{\chi}}{\Gamma_X {\rm Br}_{\chi}} \frac{d\Gamma}{d E_{\chi}} , \quad
 \int dE_{\chi}   \frac{dN}{dE_{\chi}}  = N_{\chi},
\end{align}
where $N_{\chi}$ is the multiplicity of $\chi$ and $ {\rm Br}_{\chi} $ is
the branching ratio into $\chi$ in the decay with energy-differential
width $d\Gamma/dE_{\chi}$. For a 2-body decay $X\to 2 \chi$, the
injection spectrum is a $\delta$-function ($N_{\chi} = 2$),
\begin{align}
\label{eq:delta-like}
 \left. \frac{dN}{dE_{\chi}}\right|_{\text{2-body}} = N_{\chi} \delta(E_{\chi} - E_{\rm in}), \quad E_{\rm in} = m_X/2 ,
\end{align}
 broadened only by the velocity dispersion of $X$
prior to decay; for our purposes this is a negligible effect and in
the following we take $X$ at rest. 

The galactic particle flux at earth is found from the usual line of
sight integral,
\begin{align}
    \frac{d\phi_{\mathrm{gal}}}{dE_{\chi}} =  \frac{\kappa {\rm Br}_{\chi} e^{-t_0/\tau_X} }{\tau_X m_X}
 \frac{dN}{dE_{\chi}} \times R_{\rm sol} \rho_{\rm sol} \langle J  \rangle.
\end{align}

Assuming no directional sensitivity, we average the $J$-factor over
all directions.  Its value is relatively insensitive to the employed
density profile and we use $ \langle J \rangle \simeq 2.1$ as obtained
from a NFW profile; $R_{\rm sol} = 8.33\,\rm kpc$ and
$\rho_{\rm sol} = 0.3 \,\GeV/\cm^3$ are the distance to the galactic
center and DM density at the position of the Earth.

\begin{figure}
\center
\includegraphics[width=\columnwidth]{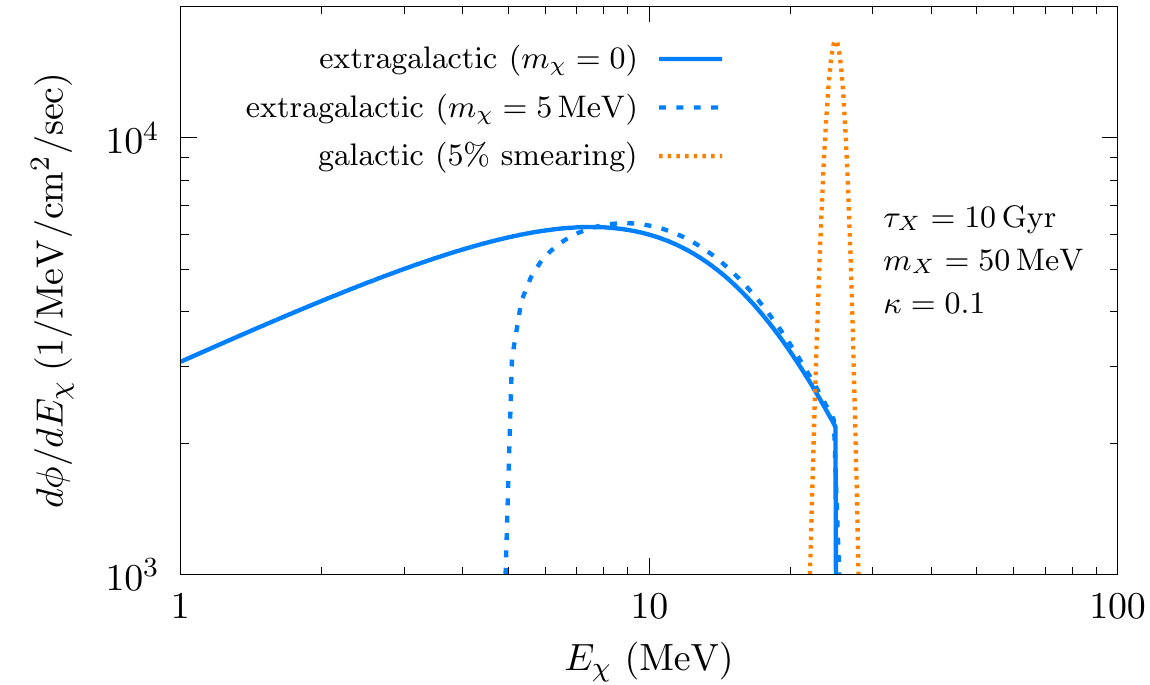}%
\caption{\small Galactic and extragalactic differential particle
  fluxes from monochromatic 2-body decay $X\to 2 \chi$. The solid
  (dashed) line is for a massless (5~MeV) daughter particle.}
\label{fluxes1}
\end{figure}

\begin{figure}
\center
\includegraphics[width=\columnwidth]{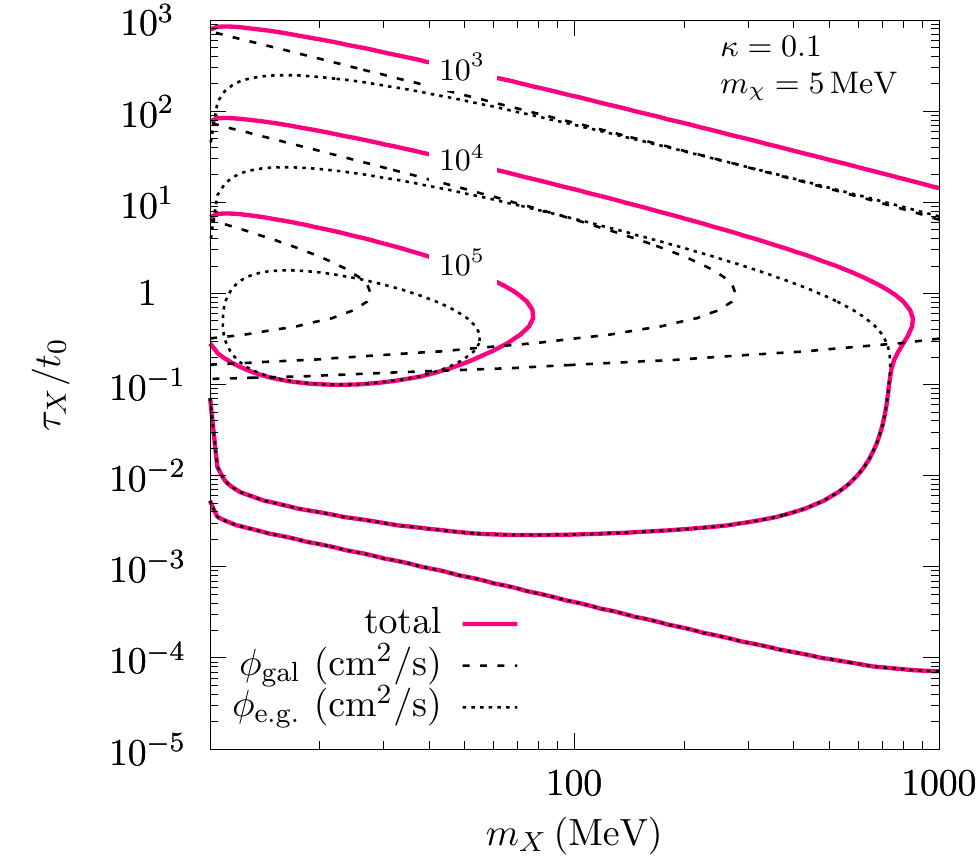}%
\caption{\small Integral galactic (``gal.'') and extragalactic
  (``e.g.'')  particle fluxes from monochromatic 2-body decay
  $X\to 2 \chi$. The labeled contours are in units of
  $1/\cm^2/\sec$. A mass of $m_{\chi} = 5$~MeV has been assumed.}
\label{fluxes2}
\end{figure}

For the extragalactic particle flux incident on Earth it is important
to take the redshift of momentum into account.
The flux, originating from the cosmological unclustered DM abundance, is
given by the redshift integral,
\begin{align}
  \label{eq:eg}
  \frac{d\phi_{\mathrm{e.g.}}}{dE_{\chi}} = 
  \frac{\kappa {\rm Br}_{\chi}  \Omega_{\rm dm}\rho_c  }{\tau_X m_X} 
\int_0^{z_f} dz\,  \frac{ e^{-t(z)/\tau_X}}{H(z)}  \frac{dN [ E_{\rm em}(z)]}{dE_{\chi}} v_{ \rm em}(z) ,
\end{align}
where the subscript ``em'' stands for the moment at emission.
Equation~(\ref{eq:eg}) reduces to the well known result of diffusive
extragalactic photon flux (in the limit of zero optic depth) when the
mass of the daughter particle vanishes, $m_{\chi}\to 0$.  The only
differences for $m_{\chi} \neq 0 $ are contained in the properly
blue-shifted energy and velocity at emission, obtained from the
blue-shifted momentum,
$ p_{\rm em}(z) = (1+z) \times p_{\chi} $, where
$p_{\chi}^2 = E_{\chi}^2 - m_{\chi}^2$ and  $p_{\rm em}^2 = E_{\rm em }^2 - m_{\chi}^2$.

The cosmological redshift information in (\ref{eq:eg}) is given
through the Hubble expansion rate
$H(z) = H_0 \sqrt{(1+z)^3 \Omega_{\rm m} + \Omega_\Lambda}$ and the
cosmic time at redshift~$z$, $t(z)$. Since we are considering DM
decays in the low-redshift Universe, it suffices to add contributions
that originate from the matter-dominated era,
$z_f < z_{\rm eq} \sim 10^4$ in~(\ref{eq:eg}). For a flat Universe one
can then use the closed expression,
\begin{align}
    t(z) = 
 \frac{1}{3 H_0 \sqrt{\Omega_\Lambda} }
\ln \left[  \frac{ \sqrt{1 + (\Omega_m/\Omega_{\Lambda})(1+z)^3 } + 1 }{\sqrt{1 + (\Omega_m/\Omega_{\Lambda})(1+z)^3 } -1} \right]
\end{align}
with  $\Omega_m = 0.315$ and $\Omega_{\Lambda} = 1 - \Omega_m$. 

At last, for a 2-body decay spectrum~(\ref{eq:delta-like}), the
redshift integral in extragalactic flux can be evaluated directly to
yield,
\begin{align}
  \left. \frac{d\phi_{\mathrm{e.g.}}}{dE_{\chi}}  \right|_{\text{2-body}}& = 
 \frac{N_{\chi} }{p_{\chi}} \frac{ \kappa {\rm Br}_{\chi}  
 \Omega_{\rm dm}\rho_c   }{\tau_X m_XH_0} \frac{e^{-t(\alpha-1)/\tau_X } }
 { \sqrt{\alpha^3 \Omega_m + \Omega_{\Lambda}} } \Theta(\alpha - 1) ,
\end{align}
where $\alpha \equiv p_{\rm in}/p_{\chi} \geq 1$. Note that the cosmic
time $t(z)$ is evaluated at redshift $z=\alpha -1$ in the
exponential. Exemplary galactic and extragalactic fluxes are shown in
Fig.~\ref{fluxes1} for $\tau_X = 10\, \Gyr$, $m_X = 50\ \MeV$ and
$\kappa = 0.1$. A Gaussian of 5\%\ width has been applied to the
galactic flux for display. 

The total flux $\phi_{\rm tot}$, integrated over the whole energy
spectrum, varies over many orders of magnitude depending on the choice
parameters. Nevertheless, one can estimate the maximum possible flux
at $\kappa \sim 0.1$, $\tau_X  = 10\, \Gyr$, while taking
$m_\chi \to 0$, and keeping $m_X$ as a free parameter:
\begin{equation}
\label{phimax}
\phi_{\rm tot}^{\rm max} \sim   \frac{10\,{\rm MeV}}{m_X}\times 10^6  \, {\rm cm}^{-2} {\rm s}^{-1}. 
\end{equation}
Completely coincidentally, the value of the DR flux may become
comparable to that of $^{8}$B solar neutrinos at $m_X \sim 10$\,MeV,
and exceed diffuse SN neutrino flux by many orders of magnitude at
$m_X \sim 50$\,MeV. Fig.~\ref{fluxes2} demonstrates an example of total fluxes for varying lifetimes and $X$ masses.

\section{Scattering and absorption signature of DR}
\label{DRsignal}

\subsection{New population of SM neutrinos}
\label{newNu}

If non-thermal DR radiation consists of SM neutrinos, we can predict
their interaction rates in  dark matter and neutrino detectors.
The coherent nuclear recoil generated by the neutral current
interaction is the easiest to treat, as it has no neutrino flavor or
helicity dependence.  The coherent neutrino-nucleus recoil cross
section is given by,
\begin{align}
  \label{eq:recoilSM}
 \frac{d\sigma}{dE_R} =  Q_W^2  \frac{G_F^2  F^2(q) m_N}{ 4\pi  } \left( 1 - \frac{m_N E_R}{2 E_{\nu}^2} \right) , 
\end{align}
where $Q_W = Z (4 s_w^2 -1) + N $ is the weak charge of the nucleus.
The scattering is essentially coherent in the number of neutrons $N$,
owing to a cancellation in charge number $Z$ since the weak angle is
$s_w^2 \simeq 0.23$.  The degree of coherence is given by Helm form
factor $F(q)$~\cite{Helm:1956zz}, evaluated at $q= \sqrt{2 m_N E_R}$. Expression
(\ref{eq:recoilSM}) is used by us to calculate the expected recoil
signal in the DM direct detection experiments.

The coherent nuclear recoil is irrelevant for generating a signal in
the most sensitive neutrino detectors. Instead, we must consider
scattering on electrons (due to both, neutral and charged currents)
and charged current scattering on nuclei. Moreover, there is a strong
dependence of the expected signal on energy, flavor and helicity of DR
neutrinos. In this paper we will assume dominance of neutrinos over
antineutrinos in DR---a possibility
discussed in Sec.~\ref{sec:complex-scalar-boson}---and, for
simplicity, we will furthermore consider a flavor-universal content of
DR at the interaction point.

From a few MeV to 1 GeV there are several approximate energy ranges
that have significant differences with respect to the expected
neutrino signal. These differences, to be discussed below, stem from a
relatively large solar neutrino flux, and from different channels of
neutrino interactions with electrons and nuclei.
\begin{enumerate} 

\item
  Below 16\,MeV. Here the neutrino signal is dominated by solar
  neutrinos, with  observed $^8$B neutrino flux,
  $\phi^{\rm obs}_{^8{\rm B}} = ( 5.16^{+0.025}_{-0.017} )\times
  10^6\,{\rm cm}^{-2}{\rm s}^{-1}$~\cite{Bergstrom:2016cbh}.  Solar
  models predict~\cite{Vinyoles:2016djt}
  $\phi_{^8{\rm B}} = 5.46(4.50) \times 10^6\,{\rm cm}^{-2}{\rm
    s}^{-1}$, depending on employed solar abundances
  \cite{Grevesse:1998bj} (\cite{Asplund:2009fu}).  Taking the
  difference in the prediction as indicative of the size of the
  systematic error bar, one can conservatively, albeit somewhat
  ad-hoc, limit any additional flux of DR neutrinos in this energy
  range at nominal 90\%~C.L.~as
  \begin{align}
    \label{eq:atm}
    \ph_\nu <1.6 \times 10^6\,{\rm cm}^{-2}{\rm s}^{-1}.
  \end{align}
  The main signature for neutrinos in this energy range is the
  scattering on electrons due to charged (CC) and neutral (NC)
  currents.

\item 16\,MeV to 30\,MeV. In this energy range the most important
  neutrino component is $\bar\nu_e$. The CC rate due to the inverse
  beta decay process on free protons, $p+\bar\nu_e\to n +e^+$ has a
  large cross section, and almost the entire anti-neutrino energy is
  communicated to the positron, $E_{e^+}\simeq E_{\bar \nu} -1.8$~MeV.
  Notice that there is no corresponding counterpart process for
  neutrinos, as $n+\nu_e\to p +e^-$ is suppressed due to the binding
  energy of neutrons in carbon or oxygen nuclei. The resulting
  electron energy is at least $\sim 15$ MeV lower than $E_\nu$, which
  puts these events in the energy range completely dominated by solar
  neutrinos. Thus, apart from $\bar\nu_e$, in this energy range all
  neutrinos are detected due to their interaction with electrons
  inside the detector.

\item 30\,MeV to 150\,MeV. Starting at $\sim $30\,MeV, reactions with
  neutrons inside nuclei (such as {\em e.g.}
  $^{16}{\rm O}+\nu_e \to {\rm ^{16}F}+e$) are no longer kinematically
  suppressed, and the energy of electrons in the final state (reduced
  relative to the incoming neutrino energy by $\sim $15\,MeV or more)
  are above the energies of electrons created by solar neutrinos.  In
  this energy range the CC cross sections of $\nu_e$ and $\bar \nu_e$
  become similar and dominate over other forms of neutrino
  interactions. Above the muon threshold, the CC production of $\mu$
  by $\nu_\mu$ also becomes possible, and it is constrained by the
  decay signature of stopped muons.
  In general, sub-Cerenkov energy muons irrespective of their origin
  represent a challenging background for searches of neutrinos of
  lower energy, as muon decays produce electrons that have sub-50 MeV
  energies. In our analysis, we account for both the CC interactions
  of $\nu_e$ and the additional backgrounds introduced by muon decays.

\item Above 150\,MeV. This is the energy range where the flux of
  atmospheric neutrinos is well measured, and found to agree with
  model predictions with $\sim20$\% accuracy~\cite{Richard:2015aua}.
  The main interactions for $\bar\nu_{e(\mu)},\, \nu_{e(\mu)}$ are the
  CC processes with nuclei.  Therefore, using the results
  of~\cite{Richard:2015aua}, one can limit any extra DR neutrino
  component. Taking it conservatively, we demand that the DR flux
  shall not exceed $\sim 1/2$ of the atmospheric neutrino flux,
  yielding
  $d\phi_{\nu_e}/dE < 0.08\times E^{-2}\,{\rm GeV\times cm}^{-2}{\rm
    s}^{-1}$. (We stop our considerations of DR at about 1 GeV, and
  throughout this range a $\phi_\nu \propto E^{-2}$ scaling holds
  reasonably well.)  A constraint of similar strength applies to the
  muon neutrino flux above $E_\nu = 300$~MeV~\cite{Richard:2015aua}.

\end{enumerate}

To summarize, the neutrino fluxes are least known directly in the energy
regions 2 and 3. To treat constraints on DR with comparable energy
range of neutrinos we require the expressions for the elastic and CC cross
sections.  The elastic scattering on electrons is given by
\begin{align}
\frac{d\sigma^{\rm el}_e}{dE_R}& =\frac{G_F^2  m_e}{ 2\pi  } \left[ (2+g_L)^2 +g_R^2\left(1-\frac{E_R}{E_\nu}\right)^2\right],\\
\frac{d\sigma^{\rm el}_{\mu,\tau}}{dE_R}& =\frac{G_F^2  m_e}{ 2\pi  } \left[ g_L^2 +g_R^2\left(1-\frac{E_R}{E_\nu}\right)^2\right],
\end{align}
where $g_L= -1+2s_w^2$ and $g_R =2 s_w^2$. Indices $e,\mu,\tau$ stand for the neutrino flavor dependence of the cross sections.
For a flavor-universal composition, one should take $\sigma^{\rm el}_{av} = \sum_{i=e,\mu,\tau}\sigma^{\rm el}_i/3$.
Because of the solar neutrino background, the most relevant quantity is the cross section that produces electrons above 
$15$\,MeV energy, $\int_{15\,{\rm MeV}} (d\sigma^{\rm el}/dE_R )dE_R$. For a 30\,MeV neutrino such flavor-averaged cross section 
is $\sigma^{\rm el}_{av}(E_R>15{\rm MeV})= 6 \times 10^{-44}\,{\rm cm}^2$. 

The CC cross section of $\bar \nu_e$ can be calculated from first
principles \cite{Strumia:2003zx}. The same cannot be said about CC
cross sections of $\nu_e$. Application to SK will require the cross
sections for the $^{16}{\rm O}+\nu_e\to {\rm ^{16}F}+e$ reaction as a
function of energy. While it has not been directly measured over the
whole energy range, one can use the results of theoretical
calculations \cite{Auerbach:2002tw,Kolbe:2002gk,Paar:2007fi}. The
cross section of a $30\,$MeV electron neutrino $\nu_e$ on oxygen is
$\simeq 1.25\times 10^{-42}\,{\rm cm}^2$ \cite{Kolbe:2002gk}, rapidly
decreasing below this energy. Various final states, such as different
nuclear levels in ${\rm ^{16}F}$ and ${\rm ^{12}B}+\alpha$ are
possible, affecting the available measurable electron energy.
We will treat this complication by assuming that the final state
electron energy is, on average, shifted by $\sim 5$~MeV below the
threshold value, $E_{e} = E_{\nu} -20\,{\rm MeV}$ and model the
distribution by a Gaussian of $7\%$ width.  The modeling of this
reaction is coarse-grained and can certainly be improved with a more
dedicated study.  Given the imperfect energy resolution of water
Cerenkov detectors, we expect that the resulting limits will only
depend mildly on our assumption.

The most important question to address now are the direct limits on DR
neutrino fluxes that can be inferred from the SK data.  As energy
regions 1 and~4 are well understood in terms of neutrino fluxes, we
need to determine the acceptable level of DR neutrino fluxes for
regions 2 and~3. The experimental data relevant for this energy range
are reported as a search for the diffuse supernova neutrino
background~\cite{Bays:2011si}. While the SK collaboration applies
their search to limit $\bar\nu_e$, owing to the fact that the CC cross
section is largest, the same data can be used to limit other neutrino
fluxes; relevant information can also be extracted from the high-end
part of the $^8$B solar neutrino spectrum~\cite{Abe:2010hy}. In the
25-to-75\,MeV range the background for the supernova $\bar\nu_e$
search is dominated by decay electrons that are produced from muon
decays---themselves sourced from $\nu_{\mu}$ but undetected in their
Cerenkov radiation---inside the SK volume, and by residual atmospheric
electron neutrino CC events. A simultaneous fit to the shape of the
signal plus background components allows the SK collaboration to
extract a tight constraint on the flux of~$\bar \nu_e$.

\begin{figure}
\center
\includegraphics[width=0.8\columnwidth]{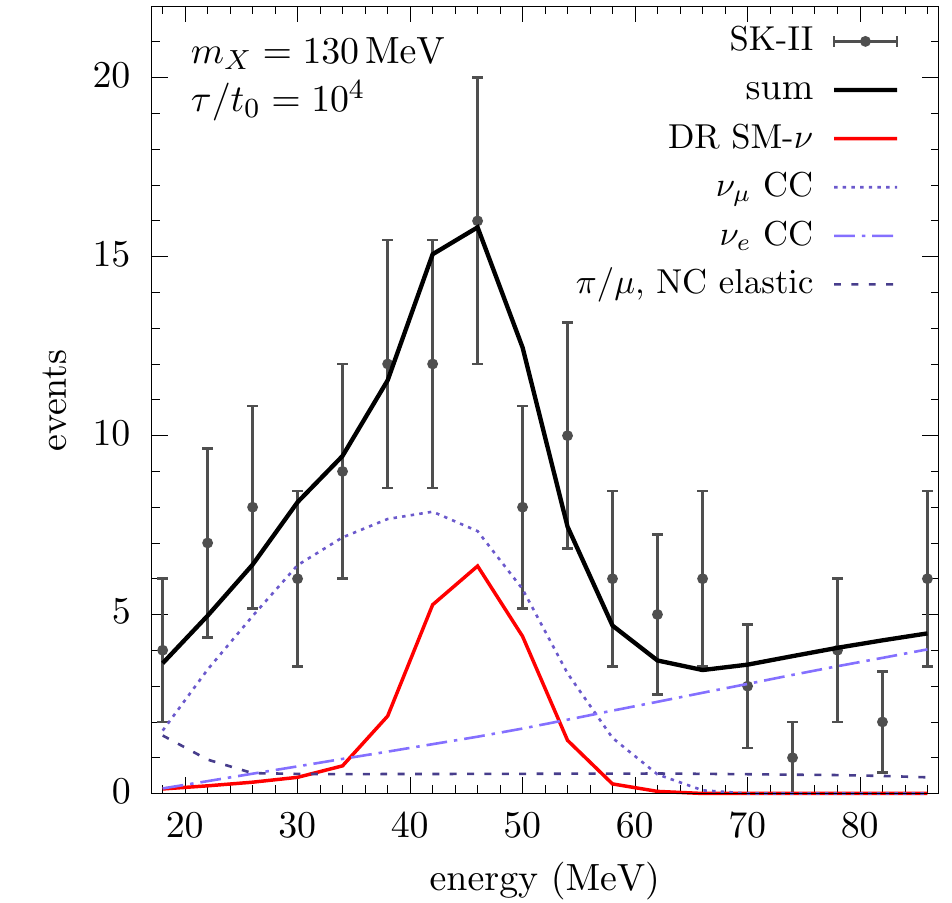}%
\caption{\small Exemplary fit to the data of the SK-II run in the SN
  neutrino search window between 18-86~MeV~\cite{Bays:2011si}. Besides
  the known, fixed background (dashed line) and the new DR-induced
  contribution (red line) for the chosen combination
  $\tau_X/t_0 = 10^4$ and $m_X = 130\,\MeV$, backgrounds of known
  shape but uncertain amplitude (dotted and dash-dotted lines) are
  determined in strength by a fit.}
\label{dsnb}
\end{figure}

We adopt a similar strategy and obtain a fit by taking into account
the above sources of background with floating normalization (in
addition to amplitude-fixed, smaller backgrounds that are inferred
from ``sidebands'' to the SK search window; see Figs.~14-16 in
\cite{Bays:2011si}), together with the DR signal calculated via its
elastic scattering on electrons and CC scattering of $\nu_e$ on oxygen
as outlined above; to the new component we apply the signal efficiency
as reported in~\cite{Bays:2011si}. For the fit we minimize the
likelihood ratio~\cite{Baker:1983tu},
\begin{align}
  \label{eq:likelihood}
  -2 \ln \lambda = 2 \sum_{i=1} \left[ \mu_i - n_i + n_i \ln \frac{n_i}{\mu_i} \right],
\end{align}
where $n_i$ ($\mu_i$) is the observed (expected) number of events
 in bin $i$ and the sum runs over all bins of
the SK search region (middle panels of Figs.~14-16 in
\cite{Bays:2011si}).
In order to keep the scope of this investigation under control, in our
numerical study we consider only the data from the SK-II period with
22.5~kton fiducial volume and 794~days of livetime; the inclusion of
other SK runs is not expected to change the results qualitatively.
An exemplary fit ($p$-value 0.71), currently allowed by the data, is
shown in Fig.~\ref{dsnb}. The red line is the DR signal, dominated by
the CC reaction on oxygen.

\begin{figure}
\center
\includegraphics[width=\columnwidth]{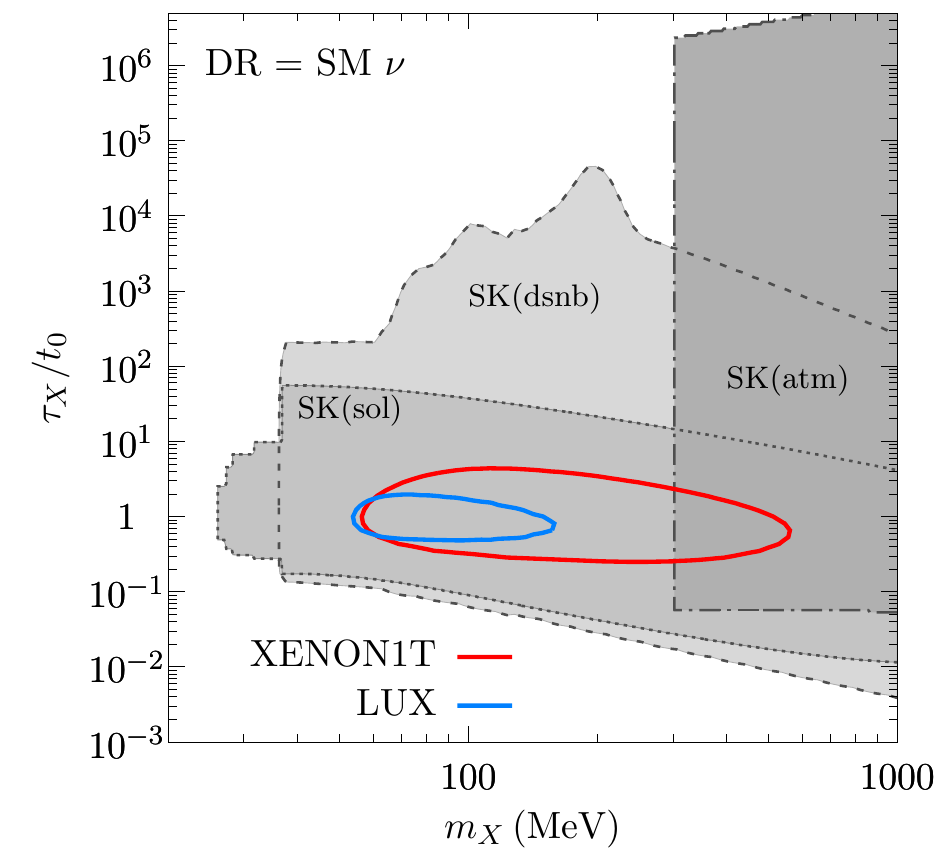}%
\caption{\small Constraints on SM neutrino DR, at 95\% C.L.,in the assumption of
  $X\to \nu\nu$ decay, with equal weight for each neutrino flavor from
  SK measurements of atmospheric (atm) and solar (sol) fluxes as well
  as from searches for a diffuse SN neutrino background (dsnb). The
  limits supersede the current direct detection constraints derived  from the
  final data set of LUX and from the initial data set of XENON1T; see
  main text for details.}
\label{constraintsSM}
\end{figure}

The resulting likelihood fit of SK-backgrounds at each point in the
$(m_X,\tau_x)$-plane produces constraints on the admissible amount of
DR. This constraint, at 95\% C.L., is shown as the gray area labeled
SK(dsnb) in Fig.~\ref{constraintsSM}, assuming that, close to its
minimum, (\ref{eq:likelihood}) follows a $\chi^2$-distribution. In
addition, the gray areas SK(atm) and SK(sol) are excluded from
atmospheric and solar flux measurements using~(\ref{eq:atm}) and
reported fluxes in~\cite{Abe:2010hy}, respectively.
The SK excluded regions supersede current constraints that are imposed
by LUX~\cite{Akerib:2016vxi} and XENON1T~\cite{Aprile:2017iyp}. We
derive the latter limits by computing the S1 scintillation signal from
probability distribution functions that we derive with the statistical
model described in~\cite{Lang:2016zhv}; see also the supplementary
material of~\cite{An:2017ojc}. As input, the mean of the S1 signal is
obtained from the light yield curve of Fig.~1 of~\cite{Akerib:2015rjg}
with a minimum nuclear recoil of 0.7~\keV\ imposed. The employed
overall light collection efficiencies are $g_1= 0.1$ and $0.144$,
respectively. The generated signals are then constrained against data
using the `maximum gap method'~\cite{Yellin:2002xd} at 95\%~C.L..

To pick a specific example, we derive the current constraint on the
neutrino flux originating from the decay of a DM particle of
$m_X=50$\,MeV and lifetime $\tau_X \gg t_0$ : \be \phi_{\nu}(E_\nu
\simeq 25\,{\rm MeV}) < 5\times 10^{2} {\rm \,cm}^{-2}{\rm s}^{-1}.
\ee Notice that this constraint is more than two orders of magnitude
more relaxed compared to the SK limit on a cosmic $\bar\nu_e$ flux.
Consequently, if this limit is saturated by DR, then the expected
scattering rate inside the xenon-based direct detection experiments
will be such as to mimic a DM particle with $m_{\rm DM} \sim 30$\,GeV
and cross section of $\sigma \sim 10^{-47}$cm$^2$, which is
significantly above the traditionally derived ``neutrino floor"; a more detailed numerical study on this point is in preparation.

The SK constraints on $\bar\nu_e$ can be significantly improved with the
addition of Gd~\cite{Beacom:2003nk}, as it would allow efficient
detection of final state neutrons. This detector modification is
unlikely to help strengthening constraints on other neutrino species,
and therefore we project that SK constraints on $\nu$ DR are unlikely
to be improved. On the other hand, the next generation of liquid xenon
DM detectors is likely to reach the $10^{-47}$cm$^2$ level of
sensitivity to DM, which would also make them efficient probes of
neutrino DR. It has to be said that {\em if} the signal is detected at
that level, one would have to check whether it is coming from
scattering of nuclei on relativistic (DR) or non-relativistic (DM)
species.  This can \textit{e.g.}~be done, in principle, if another
large and sensitive DM detector, based on a significantly lighter
target nucleus, is built to complement xenon-based experiments. In
particular, large scale argon-based detectors \cite{Aalseth:2017fik}
may fill that niche.

\subsection{New fermions interacting with SM via a dark force}
\label{newChi}

Here we consider $\chi$ interacting through a vector portal that we
take to be the baryonic current $J_B^{\mu}(\vec x)$, mediated by a
massive vector $V_{\mu}$ with mass $m_V$.
The incoming (semi-)relativistic particles $\chi$ will induce elastic
scattering on nuclei in direct detection experiments. 
For that reason, we will generalize (\ref{eq:recoilSM}) 
to be valid both in the relativistic
($E_{\chi} \geq m_{\chi}$) and non-relativistic limits
($E_{\chi} \leq m_{\chi}$) alike.
In order to compute the elastic recoil cross section $d\sigma / dE_R$
we note that the manifestly spin-independent (SI) part in the
$\chi$-nucleon matrix element
$ \langle \vec p_n' | J^{\mu}_B(0) | \vec p_n \rangle = \bar u_{\vec
  p_n'} \Gamma^{\mu} u_{\vec p_n}$ is given by the vertex factor
\begin{align}
  \Gamma_{\rm SI}^{\mu}(q) = 2 m_n \left[F_b(q) - \frac{q^2}{4m_n^2} F_m(q)\right] \frac{ (p_n
  + p_n' )^{\mu}}{(p_n + p_n' )^2} ,
\end{align}
where $F_b(0) = 1$ is the baryon number of the nucleon and in the
following we can take it to unity; furthermore, we drop the term
proportional to $F_m$ since it is suppressed. 
The direct detection recoil cross section on nuclei is then found to be
\begin{align}
  \label{eq:recoilNub}
   \frac{d\sigma}{dE_R} = A^2 \frac{8\pi \alpha_B^2 Q_\chi^2 F^2(q) m_N}{ (m_V^2 + 2m_N E_R)^2  }
   \left[ 1 + \frac{m_{\chi}^2}{ |\vec p_{\chi}|^2  } - \frac{m_N E_R}{2 |\vec p_{\chi}|^2  }\right] . 
\end{align}
The nuclear form factor $F(q)$ is the same as above in (\ref{eq:recoilSM}).
For $m_{\chi} \to 0$, $|\vec p_{\chi}| \to E_{\chi}$ and
Eq.~(\ref{eq:recoilNub}) reduces to the expression obtained
in~\cite{Pospelov:2012gm} using $G_B \equiv Q_\chi g_B^2/m_V^2$ and when the
interaction obeys the contact limit $q^2 \ll m_V^2$. 

\begin{figure}
\center
\includegraphics[width=\columnwidth]{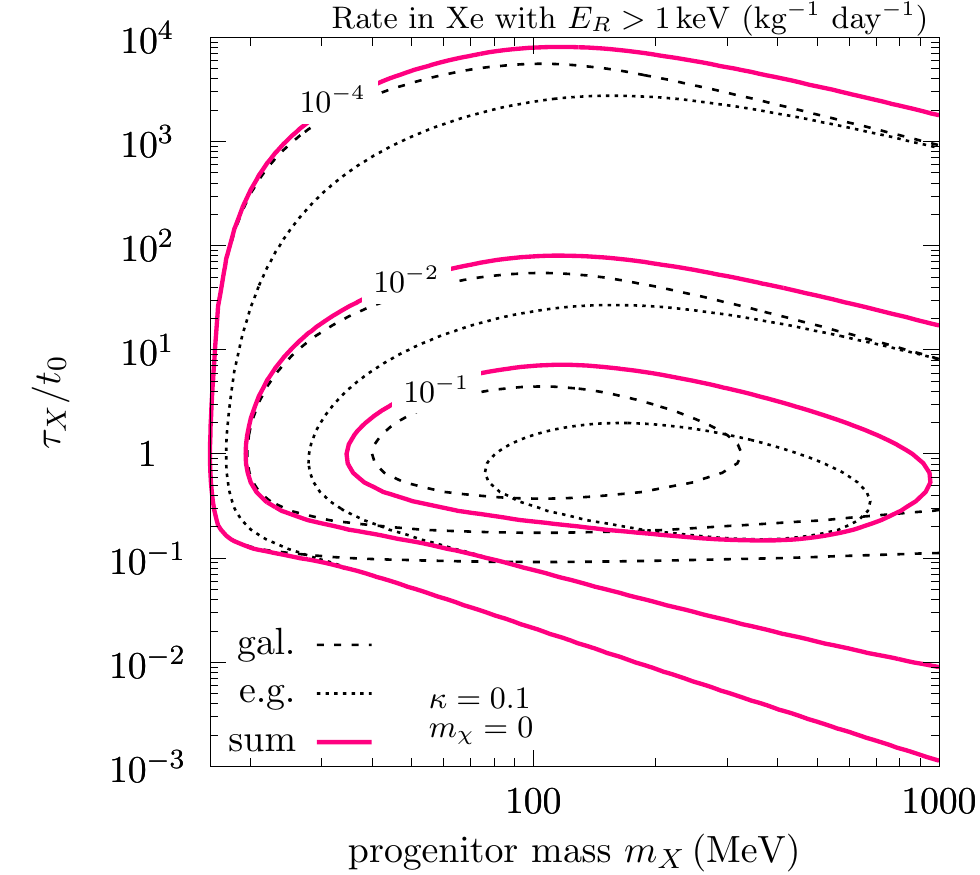}%
\caption{\small Predicted recoil rate $R(E_R>1\,\keV)$ for DR
  consisting of massless $\chi$ particles that are coupled to the
  baryon current with strength $G_B = 10G_F$, in units of
  events/kg/day, and for xenon as detector material.}
\label{dR}
\end{figure}

\begin{figure}
\center
\includegraphics[width=\columnwidth]{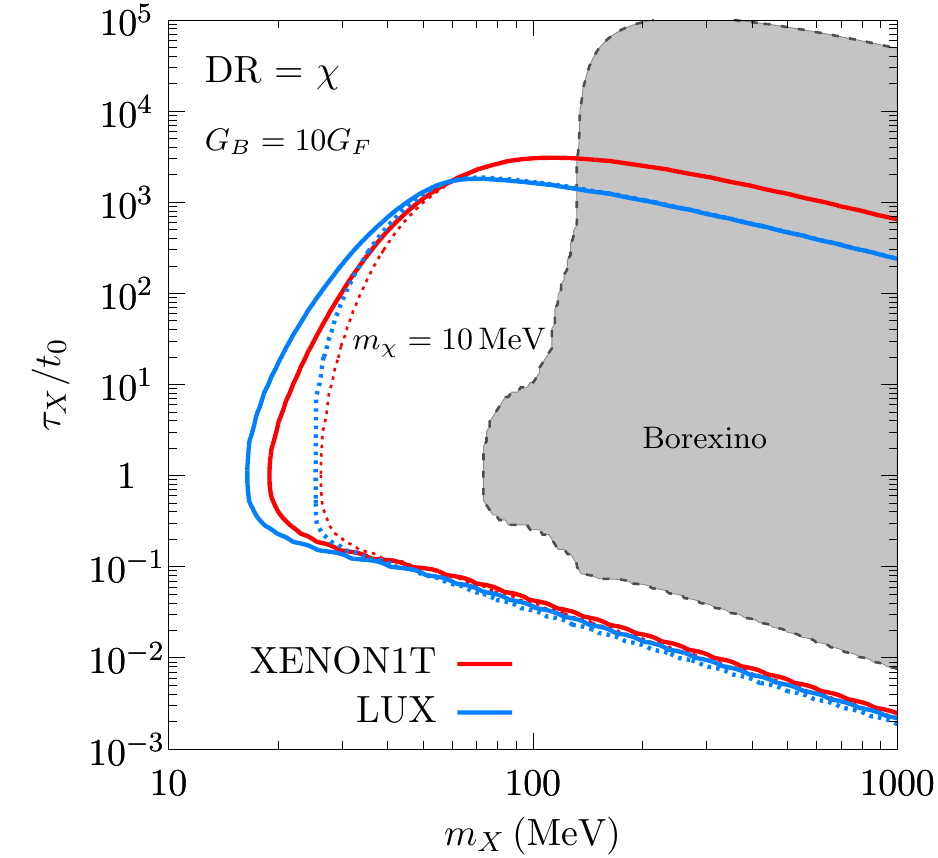}%
\caption{\small Constraints on DR coupled to SM, at 95\% C.L., through gauged baryon
  number with $G_B = 10G_F$ from the direct detection experiments LUX
  and XENON1T as well as from the neutrino experiment Borexino. Two
  cases are shown: the solid line is for $m_\chi=0$ and dotted line is for
  $m_\chi = 10$\,MeV; for Borexino, the difference in assumed mass is not visible.}
\label{constraints}
\end{figure}

The recoil rate in a direct detection experiment is then given by,
\begin{align}
  \frac{dR}{dE_R} =  N_T  \sum_{i=\rm gal, eg} \int_{E_{\chi, \rm min}}^{E_{\rm in}} dE_{\chi}\, 
\frac{d\phi_i}{dE_{\chi}} \frac{d\sigma(E_{\chi})}{dE_R} . 
\end{align}
where $N_T$ is the target number density per detector mass (for
composite materials the rate is to be summed over the various
target elements).
The minimum $\chi$ energy $E_{\chi, \rm min}$ to produce a recoil $E_R$
is obtained from the corresponding minimum momentum
$ |\vec p_{\chi, \rm min}|  = \sqrt{E_R m_N/2} $.
The predicted recoil rate in liquid xenon detectors is shown in
Fig.~\ref{dR}. Current constraints from LUX~\cite{Akerib:2016vxi} and
XENON1T~\cite{Aprile:2017iyp} on the model for $G_B = 10\,G_F$ is
shown in Fig.~\ref{constraints}. The solid (dotted) lines are for $m_{\chi} = 0$
(10~MeV); for details on the procedure that goes into the derivation
of these limits cf.~the preceding section.

To include constraints on baryonic-current coupled DR in $\chi$ from
neutrino experiments, we have to consider the scattering of $\chi$ on
nuclei that may lead to their recoil, break-up or excitation. We refer
to previous studies~\cite{Pospelov:2012gm}, where it was shown that in
the limit of small momentum transfer (in units of the nuclear size
$R_N$), the interaction rate for inelastic processes is suppressed by
$(qR_N)^4$, which is a very small factor for $q \ll 100$\,MeV. This
allows to tolerate large values of $G_B$, without running into strong
neutrino constraints. For this paper, we include constraints imposed
on the model by scintillator-based neutrino detectors such as Borexino
\cite{Alimonti:2008gc}, that result from the elastic scattering on
protons.  The scattering of $\chi$ on protons, depending on $E_\chi$,
can give a significant proton recoil that leads to energy deposition
inside a liquid scintillator, and will give significant constraints on
the higher end of the energy range of DR considered in this paper.

To calculate the proton recoil, we use (\ref{eq:recoilNub}) but
take the nuclear form factor  to be of the dipole form,
$F_b(q^2) = (1+q^2/(0.71\,{\rm GeV}^2))^{-2}$~\cite{Perdrisat:2006hj}.
The treatment of the quenching factor in the recoil of protons is
obtained following~\cite{Dasgupta:2011wg} and references therein; see
also~\cite{Pospelov:2012gm}. The gray region in Fig.~\ref{constraints}
shows the resulting excluded region from measurements by Borexino, and
it is derived from the fiducialized data shown in Fig.~3
of~\cite{Bellini:2012kz}.

\subsection{Discussion of ALPs and dark photons as DR}
\label{newAlps}

Dark radiation in the form of ALPs or dark photons can appear as a
result of $X\to aa$ or $X \to VV$ decays%
.  The main question to investigate here is whether DM decay can
provide a flux of ALPs or dark photons in the range where they can be
detected while respecting limits from other, primarily astrophysical,
constraints. In the case of ALPs, the most explored range is the keV
frequency range, which for our scenario would imply
$ m_X \geq O({\rm keV})$, and the interaction with photons,
${\cal L}_{\rm int} = g_{a\gamma\gamma}a {\vec E\cdot \vec B}$. This
energy range could make both, the dark matter experiments as well as
axion helioscopes, sensitive to ALP DR.  Previous studies have
concluded that ALP DM with $m_a$ in the keV range is already severely
restricted by the absence of a serious excess in $X$-rays (for a
recent summary of X-ray constraints on decaying dark matter see
\textit{e.g.}~\cite{Abazajian:2017tcc}). The indirect constraints on
ALP DM are typically much stronger than those provided by direct
searches \cite{Pospelov:2008jk}. To remove constraints resulting from the
decay of ALPs, one needs to require an additional hierarchy,
$m_a\ll m_X$.

The easiest way to assess the detectability of DR in form of ALPs is
to compare their maximal flux with the flux of solar axions.  The
spectral flux of the latter is given in~\cite{Andriamonje:2007ew}, and
integrating it over energy, one obtains
$\phi_a^{\rm solar} \simeq 10^{12} (g_{a\gamma\gamma}\times
10^{10}\,{\rm GeV})^2{\rm cm}^{-2}{\rm s}^{-1}$. Comparing
$\phi_a^{\rm solar}$ to the maximum flux of ALPs attainable through DM
decay, $\phi_{\rm tot}^{\rm max}$ in (\ref{phimax}), we arrive at the
{\em maximum} value of coupling $g_{a\gamma\gamma}$ when the DR flux
of ALPs have a chance of becoming larger than the solar flux,
\begin{equation}
g_{a\gamma\gamma} \leq 10^{-11}\,{\rm GeV}^{-1} \times \left( \frac{1\,{\rm keV}}{m_a}\right)^{1/2}.
\label{benchmark}
\end{equation} 

Current solar helioscopes utilize the $a\to \gamma$ conversion in the magnetic field to search for solar ALPs. The same techniques can be 
used to search for ALPs forming DR. (For some values of parameters, the galactic component of DR dominates, and one should expect an enhancement of the conversion in the direction to the galactic center.)

It is easy to see that in the keV range of frequencies, the benchmark
value $g_{a\gamma \gamma} = 10^{-11}\,{\rm GeV}^{-1} $ in
Eq.~(\ref{benchmark}) is outside the reach of the current generation
keV-range ALP detector CAST~\cite{Andriamonje:2007ew}, but may be
amenable to searches with the next generation ALP telescopes, such as
IAXO~\cite{Giannotti:2016drd}.  A similar conclusion can be reached
for ALPs coupled to the electron spin via
$g_{aee} a \times \bar e\gamma_5e$.  The expected solar flux of ALPs
is at the level of
$ \sim(g_{aee} \times 10^{13})^2 \times 10^9 {\rm cm}^{-2}{\rm
  s}^{-1}$ \cite{Redondo:2013wwa}, which is again somewhat larger than
the maximum attainable flux for DR with a keV scale progenitor $X$,
unless $g_{aee}$ is below $10^{-13}$. The current sensitivity of dark
matter experiments to solar axions is at the level of
$g_{aee}\sim 8\times 10^{-12}$ \cite{Aprile:2014eoa}, and, therefore,
only significant improvements in the sensitivity of large-scale dark
matter experiments could render a hypothetical ALP component of DR
detectable.

We now turn to the case of dark radiation in the form of dark
photons. The main difference with the ALP case is that for small mass
of dark photons, the solar flux decouples as
$\phi_V^{\rm sol} \propto m_V^2$ \cite{An:2013yfc}. (See
Ref.~\cite{Redondo:2015iea} for a detailed calculation of the solar
energy loss to dark photons.) On the other hand, the production of DR
dark photons may not need to be suppressed by small mass $m_V$ in the
same limit, and therefore the flux of DR can be parametrically larger
than the solar flux of dark photons. The analysis of the absorption of
dark photon dark radiation is very similar to the analysis performed
for dark photon DM \cite{An:2014twa,Bloch:2016sjj}. The conclusion of
these studies is that in some corners of mass--mixing angle parameter
space, $\{m_V, \epsilon \}$, the direct detection experiments have
sensitivity to dark photons beyond the astrophysical
constraints. Unfortunately, our analysis shows that the dark photon
dark radiation is currently not constrained by direct detection
experiments. The qualitative reason for that is as follows: when the
medium effects can be neglected ($m_V > 1$\,eV) the absorption rate
per atom scales as
$\epsilon^2 \times n_V \times \sigma_{\rm photo} \times c$, where
$n_V $ is the number density of dark photons (in the form of DM or
DR), $\sigma_{\rm photo}$ is the photo-ionization cross section, and
$c$ is the speed of light. The rate is approximately independent on
the velocity of dark photons when their total energy is fixed.  For
dark photon DM the number density $n_V$ is typically larger than for
dark photon DR, resulting in a smaller ionization rate for the latter.
Taking a representative point on the parameter space,
$\epsilon = 10^{-11}$, $m_V = 1$\,eV, and $\kappa= 0.1$, $m_X = 50$
eV, and $\tau_X = t_0$, we find that the ionization rate for the
XENON10 experiment is not exceeding a few times $10^{-7}$/kg/day,
which is much smaller than the current sensitivity.

Finally, if the dark photon mass is very small or zero, it can mediate
the interaction between charged particles of the SM and the dark
sector $\chi$ particles when they are charged under the dark photon
gauge symmetry. Such objects, generically called milli-charged
particles, have been extensively studied in the literature with the
latest constraints compiled in \cite{Vogel:2013raa}. While the range
of small masses is again generically very constrained by the
combination of cosmology and astrophysics, the GeV and heavier range
$\chi$ are allowed to have charge up to $(10^{-2}-10^{-1})\times e$.
If the DM particle $X$ decays to a pair such millicharged particles,
one should expect a variety of new effects associated with scattering
of $\chi$ inside dark matter and neutrino detectors.

\section{Conclusions}
\label{DC}

We have considered a hypothetical possibility that along with
non-relativistic DM, some (semi-)relativistic particles form a cosmic
dark radiation (DR) background that may have a noticeable interaction
rate with the SM particles. Such DR can be a non-thermal component of
energetic neutrinos, or SM singlet particles. The most efficient
mechanism for populating DR radiation states is the decay of DM, and
if it happens at late redshift, a significant fraction of DM is
allowed to decay to DR. Adopting this framework, we have derived the energy spectrum of DR
that includes the galactic and a global cosmological component, as a
function of the progenitor particle's lifetime, mass and abundance.

To narrow the discussion we have concentrated on the sub-GeV range for
the energy of DR (or, equivalently, the mass of decaying DM
particles).  This range is the most relevant one for a potential
signal in the detectors that are built to search for WIMP DM recoils.
Therefore, they could also register energy and momentum transfer that
is communicated to nuclei and electrons via their interaction with DR.
For the masses of decaying particles in the tens of MeV range, the
resulting fluxes of DR particles may reach
$10^5-10^6\,{\rm cm}^{-2}{\rm s}^{-1}$, which is a fairly significant
flux, exceeding atmospheric and expected diffuse SN neutrino fluxes by
many orders of magnitude. Correspondingly, we find that both, neutrino
and DM direct detection experiments are sensitive to the weak-scale
interaction of DR with  SM fermions (nucleons and electrons).

For DR in the form of the SM neutrinos, we find that the
Super-Kamiokande (SK) experiment provides the dominant
constraints. The strongest constraints stem from limits on
$\bar \nu_e$ fluxes. If, however, DM decays preferentially to
neutrinos, rather than antineutrinos, the constraints become much
milder, and current SK data can tolerate much larger fluxes of DR
neutrinos. If one saturates our derived limits on DR neutrino fluxes,
DM direct detection experiments need to improve their sensitivity by
approximately two orders of magnitude to become competitive with SK.
Another way of phrasing the same finding is to say that the signal
from DR-induced neutrino-nucleus scattering could be significantly
above the normally expected ``neutrino floor" for DM direct detection
experiments. In particular, we find that for the energy of DR
particles in the $\sim 25 $~MeV range, the gain over the nominal
neutrino floor can be very significant, providing a potential signal
that would compare to a one from a $\sim 30 $ GeV WIMP-type dark
matter particle with a scattering cross section of
$10^{-47}\, {\rm cm}^2$.

Among the most interesting cases for DR in the New Physics sector are
the dark radiation models that interact with nuclei via a baryonic
current, which is a possibility that is least constrained by neutrino
experiments. Here, the existing direct detection experiments provide
dominant bounds for the same DR energy range.  Dedicated studies with
upcoming neutrino experiments such as SNO+, JUNO, and DUNE/LBNF can
potentially be complementary probes for these models.

To conclude, the co-existence of DM and interacting DR is a generic
possibility that draws an analogy with the structure of the SM, where
both massive particles (atoms) and radiation (photons, neutrinos) are
present. This work demonstrates that this broad new class of physics can be probed with experiments originally designed to search  for dark matter and study neutrino interactions.

\section*{Acknowledgments}

We would like to thank M.~Nikolic and J.~Ruderman for useful discussions.  Research
at Perimeter Institute is supported in part by the Government of
Canada through NSERC and by the Province of Ontario through MEDT. JP
is supported through the New Frontiers Program by the Austrian Academy
of Sciences.

\vfil\pagebreak

\bibliography{Refs}
\end{document}